\documentclass[pre,aps,showpacs,superscriptaddress,floatfix]{revtex4-2}

\usepackage{mathtools}
\usepackage[usenames,dvipsnames]{xcolor}
\usepackage{amsmath}
\usepackage{amssymb,mathrsfs}
\usepackage{graphicx}
\usepackage[colorlinks=true]{hyperref}
\usepackage{soul}
\usepackage{xcolor}

\usepackage{float}

\newcommand{\prt}{\partial}
\newcommand{\al}{\alpha}

\newcommand{\om}{\omega}

\newcommand{\vphi}{\varphi}

\newcommand{\rt}{\widetilde{\rho}}

\begin{document}

\title{Hamiltonian mechanics of ``magnetic'' solitons in two-component
Bose-Einstein condensates}

\author{A.~M.~Kamchatnov}
\affiliation{Institute of Spectroscopy, Russian Academy of Sciences, Troitsk, Moscow, 108840, Russia}
\affiliation{Skolkovo Institute of Science and Technology, Skolkovo, Moscow, 143026, Russia}

\begin{abstract}
We consider motion of a ``magnetic'' soliton in two-component condensates along a non-uniform
and time-dependent backgrounds in framework of the Hamiltonian mechanics. Our approach is based on
generalization of Stokes' remark that soliton's velocity is related with its inverse half-width
by the dispersion law for linear waves continued to the region of complex wave numbers. We
obtain expressions for the canonical momentum and the Hamiltonian as functions of soliton's
velocity and transform the Hamilton equations to the Newton-like equation. The theory is
illustrated by several examples of concrete soliton's dynamics.
\end{abstract}



\maketitle

{\it Dedicated to the memory of David Kaup}

\section{Introduction}

The idea that solitons behave under action of smooth enough external forces similar to
point-like particles of classical mechanics is well known. In particular, a number of
perturbation methods is based on this qualitative picture \cite{kaup-76,km-77,km-78,ka-81}
(see also Refs.~\cite{km-89,og-2000} and references therein) and these theories are confirmed
by numerical and real experiments. The standard perturbation theory is quite involved in
case of dark solitons propagating along a non-uniform background \cite{ky-94,ba-2000,anhf-11}
when separation of the soliton's dynamics from dynamics of the surrounding background
is not obvious: a moving soliton generates a `shelf' in the background and this leads to an
essential back reaction on the soliton's motion. For example, the frequency of oscillations
of a dark soliton in a Bose-Einstein condensate (BEC) confined in a harmonic trap with the
frequency $\om_0$ is equal to $\om_0/\sqrt{2}$ whereas the center of gravity of BEC still
oscillates with the trap frequency \cite{ba-2000}. This means that the soliton's motion
generates a counterflow in the surrounding condensate which results in an essential
redistribution of its mass. In this sense, motion of dark solitons can be likened \cite{kp-04}
to motion of quasi-particles in solid-state physics where dynamical properties of
quasi-particles are determined by their interaction with the surrounding matter. As a result, 
derivation of equations of motion of such solitonic quasi-particles by the standard 
perturbation theory is, generally speaking, quite a difficult problem.

A simpler method has been recently suggested in Ref.~\cite{ks-23a}. It is based on the
assumption that if the nonlinear wave equation is Hamiltonian, then its reduction to the
case of the soliton's motion along a smooth time-dependent background remains also Hamiltonian.
The condition that the soliton's dynamics is Hamiltonian, combined with old Stokes' remark
\cite{stokes} that the soliton's velocity is related with its inverse half-width by the
dispersion relation for linear waves, allows one to find expressions for the canonical momentum
and the Hamiltonian of the point-like soliton. The method was applied to the KdV \cite{ks-23a},
NLS \cite{ik-22,kamch-23}, DNLS \cite{ks-24} equations and analytical results were confirmed by
numerical solutions for concrete situations.

In this paper, we will obtain the equations of motion for so-called ``magnetic'' solitons
in a two-component BEC. Such soliton solutions were obtained in Ref.~\cite{qps-16} in the
limit of small difference between the inter- and intra-components interaction. In this case,
the BEC components move in opposite directions without transfer of the total density,
so the dynamics is completely determined by the counterflow of the BEC components. We
obtain the expressions for the canonical momentum and the Hamiltonian of the soliton
dynamics and transform the Hamilton equations to the Newton-like equation for soliton's
motion. Since the dynamics is determined by the gradients of the background flow, we
assume that the Newton equation remains valid when these gradients are resulted from
the background dynamics with account of external potentials. The theory is illustrated
by concrete examples of the magnetic soliton dynamics.

\section{Basic equations}

A two-component condensate is described by the system of Gross-Pitaevskii equations
\begin{equation}\label{eq1}
  \begin{split}
  & i\psi_{1,t}+\frac12\psi_{1,xx}-(\al_1|\psi_1|^2+\al_2|\psi_2|^2)\psi_1=V_1\psi_1,\\
  & i\psi_{2,t}+\frac12\psi_{2,xx}-(\al_2|\psi_1|^2+\al_1|\psi_2|^2)\psi_2=V_2\psi_2,
  \end{split}
\end{equation}
where $\psi_{1,2}$ are wave functions of two components, $\al_1$ is the constant of
interaction between atoms in each component, and $\al_2$ is the constant of interaction
between atoms which belong to different components. We assume that
\begin{equation}\label{eq1b}
  \al_1>\al_2>0,
\end{equation}
what corresponds to the repulsive interaction between atoms and the BEC is a mixture of
the components which do not separate from each other in space (see, e.g., \cite{ps-16}).
$V_1$ and $V_2$ are the external (trap) potentials acting on the first and the second
component, correspondingly.

It is convenient to introduce pseudospin variables (see, e.g., \cite{mueller})
\begin{equation}\label{eq2}
    \left(
            \begin{array}{c}
              \psi_1 \\
              \psi_2 \\
            \end{array}
          \right)=
          \sqrt{\rho}\, e^{i\Phi/2}
          \left(
            \begin{array}{c}
              \cos\frac{\theta}2\,e^{-i\phi/2}\,e^{-i\mu_1t} \\
              \sin\frac{\theta}2\,e^{i\phi/2}\,e^{-i\mu_2t}  \\
            \end{array}
          \right),
\end{equation}
where $\rho_1=\rho\cos^2(\theta/2),\rho_2=\rho\sin^2(\theta/2)$ are the densities of the
components, so that $\rho=\rho_1+\rho_2$ is the total density of BEC. The functions
$\Phi$ and $\phi$ play the roles of potentials of velocities of the mutual and relative
motions of the components; consequently the components' velocities are equal to
\begin{equation}\label{eq2a}
  v_1=\frac12(\Phi_x-\phi_x),\qquad v_2=\frac12(\Phi_x+\phi_x).
\end{equation}
In the Thomas-Fermi approximations, Eqs.~(\ref{eq1}) give the stationary distributions of
the densities of the components confined in a trap with the potentials $V_1$ and $V_2$,
\begin{equation}\label{eq2b}
  \begin{split}
  \rho_1=\rho_{1,0}(x)=\frac{\al_1(\mu_1-V_1(x))-\al_2(\mu_2-V_2(x))}{\al_1^2-\al_2^2}),\\
  \rho_2=\rho_{2,0}(x)=\frac{\al_1(\mu_2-V_2(x))-\al_2(\mu_1-V_1(x))}{\al_1^2-\al_2^2};
  \end{split}
\end{equation}
then the total density is equal to
\begin{equation}\label{eq4}
  \rho=\rho_0(x)=\frac{1}{\al_1+\al_2}(\mu-V_1(x)-V_2(x)),
\end{equation}
where $\mu=\mu_1+\mu_2$.

Substitution of Eqs.~(\ref{eq2}) into Gross-Pitaevskii Equations (\ref{eq1}) yields
\cite{ckp-16}
\begin{equation}\label{eq3}
\begin{split}
&  \rho_t+\frac{1}{2}\left[\rho(\Phi_x-\cos\theta\phi_x)\right]_x=0,\\
&  \Phi_t+\frac{\rho_x^2}{4\rho^2}-\frac{\rho_{xx}}{2\rho}
-\frac{\cot\theta}{2\rho}(\rho\,\theta_x)_x
+\frac{1}{4}(\Phi_x^2+\phi_x^2+\theta_x^2)
+(\al_1+\al_2)\rho=-(V_1+V_2),\\
& \theta_t+\frac{1}{2 \rho}(\phi_x \rho\sin\theta)_x+
\frac{1}{2}\Phi_x\theta_x=0,\\
& \phi_t-\frac{1}{2\rho\sin\theta}(\rho\,\theta_x)_x
+\frac{1}{2}\Phi_x\phi_x-(\al_1-\al_2)\rho\cos\theta=V_1-V_2.
\end{split}
\end{equation}
We are interested in dynamics of ``magnetic'' solitons of relative motion of the components
\cite{ss-02} and this solitons correspond to fast changes of the relative density (that is
of the $\theta$-variable) and of the relative velocity $v=v_2-v_1=\phi_x$. The characteristic
width $\kappa$ of such a soliton can be easily estimated from the last equation,
($(\rho_0\kappa^2)^{-1}\sim(\al_1-\al_2)\rho$), that is in our non-dimensional variables
\begin{equation}\label{eq3a}
  \kappa\sim(\rho_0\delta g)^{-1/2},
\end{equation}
where we introduced the notation $\delta g=\al_1-\al_2$. Of course, it is supposed that the
soliton's width is much smaller than the characteristic size of distributions (\ref{eq2b})
and (\ref{eq4}). The second equation (\ref{eq3}) can be written with account of Eq.~(\ref{eq4})
in the form
$$
\Big\{\frac{\rho_X^2}{4\rho^2}-\frac{\rho_{XX}}{2\rho}
-\frac{\cot\theta}{2\rho}(\rho\,\theta_X)_X
+\frac{1}{4}(\Phi_X^2+\phi_X^2+\theta_X^2)\Big\}\kappa^{-2}=2g(\rho_0-\rho),
$$
where $2g=\al_1+\al_2$ and $X=x/\kappa$, that is $X\sim1$ within the soliton's width. The 
expression inside the curly brackets is of order of magnitude of unity and it is multiplied 
by the factor $\kappa^{-2}\sim\delta g$ which is very small if the difference $\delta g$ of
the interaction constants is small. Consequently, in this limit discussed in
Ref.~\cite{qps-16}, the total density can be considered as constant within the
soliton's width: there is no transfer of total mass due to propagation of a magnetic
soliton, i.e. $\rho=\rho_0$. Thus, we arrive at the picture of point-like soliton propagating
along smooth distributions (\ref{eq2b}) and (\ref{eq4}).

If the potentials are stationary, then $\rho_t=0$, so the first Eq.~(\ref{eq3}) gives
$\rho(\Phi_x-\cos\theta\phi_x)=j=\mathrm{const}$. We will confine ourselves to
situations when there is no flow of the total density, $j=0$, so
\begin{equation}\label{eq5}
  \Phi_x=\cos\theta\phi_x,
\end{equation}
and the remaining Eqs.~(\ref{eq3}) reduce to
\begin{equation}\label{eq6}
  \begin{split}
  & \theta_t+\theta_x\phi_x\cos\theta+\frac12\sin\theta\phi_{xx}=
  -\frac{\rho_{0,x}}{2\rho_0}{\phi_x}{\sin\theta},\\
  & \phi_t-\delta g\,\rho_0\cos\theta+\frac12\cos\theta\phi_x^2
  -\frac{\theta_{xx}}{2\sin\theta}=V_1-V_2+\frac{\rho_{0,x}}{2\rho_0}\frac{\theta_x}{\sin\theta}.
  \end{split}
\end{equation}
Here $\theta$ and $\phi$ change at distances of order of magnitude of soliton's width,
i.e., $\rho_{0,x}/\rho_0\ll \theta_x,\phi_x$ and terms with $\rho_{0,x}/\rho_0$ can be neglected.
Besides that, to simplify notation and comparison with other publications, it is conveniens
to introduce a new space coordinate $x'=\sqrt{2}x$. Then we get
\begin{equation}\label{eq7}
  \begin{split}
  & \theta_t+2\theta_x\phi_x\cos\theta+\sin\theta\phi_{xx}=0,\\
  & \phi_t-(\delta g\,\rho_0-\phi_x^2)\cos\theta
  -\frac{\theta_{xx}}{\sin\theta}=V_1-V_2,
  \end{split}
\end{equation}
where we still write $x$ instead of $x'$. The potentials $V_1,V_2,$ and the total density $\rho_0$
are slow functions of $x$. These equations can be obtained
from the Hamilton principle with the Lagrangian density
\begin{equation}\label{eq8}
  \mathcal{L}=\cos\theta\phi_t-\frac12\left[\theta_x^2-
  \left(\delta g\,\rho_0-\phi_x^2\right)\sin^2\theta\right]-(V_1-V_2)\cos\theta,
\end{equation}
where $V_1,V_2$, and $\rho_0$ are assumed constant. It coincides with the Lagrangian of a
magnetic material with an easy-plane anisotropy (see, e.g., \cite{kik-90}). If we denote
\begin{equation}\label{eq9}
  \cos\theta=w,\quad \phi_x=v,\quad \rt=2\delta g\,\rho_0
\end{equation}
then we cast Eqs.~(\ref{eq7}) to the form \cite{ckp-16,ih-17}
\begin{equation}\label{eq10}
  \begin{split}
  & w_t-\left[v(1-w^2)\right]_x=0,\\
  & v_t-\left[w(\rt-v^2)\right]_x+
  \left[\frac{1}{\sqrt{1-w^2}}\left(\frac{w_x}{\sqrt{1-w^2}}\right)_x\right]_x=V_{1,x}-V_{2,x}.
  \end{split}
\end{equation}
These are nonlinear equations for polarization waves propagating along the non-uniform
stationary background with the total density distribution (\ref{eq4}); Eqs.~(\ref{eq2b})
determine the smooth background distribution of the variable $w$. Our task is to reduce
Eqs.~(\ref{eq10}) to equations of motion of a point-like magnetic soliton.

\section{Dispersion relation of linear waves, dispersionless limit and soliton solution}

Let a small-amplitude harmonic wave propagate along a uniform condensate with constant
values of $w=w_0, v=v_0$ and zero external potentials $V_1=V_2=0$. We linearize Eqs.~(\ref{eq10})
with respect to small deviations $w'=w-w_0, v'=v-v_0$ from this uniform state and obtain
\begin{equation}\label{eq15k}
  \begin{split}
  & w'_t+2v_0w_0w'_x-(1-w_0^2)v'_x=0,\\
  &v'_t-(\rt-v_0^2)w'_x+\frac{w'_{xxx}}{1-w_0^2}+2v_0w_0v'_x=0.
  \end{split}
\end{equation}
This system readily gives the dispersion relation for harmonic waves $w',v'\propto\exp[i(kx-\om t)]$:
\begin{equation}\label{eq16k}
  \om(k)=k\left(2v_0w_0\pm\sqrt{(\rt-v_0^2)(1-w_0^2)+k^2}\right).
\end{equation}
By definition we have $|w_0|<1$, so the uniform state is modulationally unstable for large
enough values of the relative velocity, $|v_0|>\sqrt{\rt}$, of the components. We will
confine ourselves to the modulationally stable situations with $|v_0|<\sqrt{\rt}$.

Even for zero potentials $V_1=V_2=0$ the background state can be non-uniform and time-dependent,
if a large-scale wave propagates through BEC, so that the wave variables are give by the functions
$w=w(x,t)$ and $v=v(x,t)$. If the characteristic size of these distributions is much greater
than the soliton's width $\sim\kappa^{-1}$, then we can neglect the dispersion term in the second
equation (\ref{eq10}) and arrive at the equations of the dispersionless (hydrodynamic)
approximation
\begin{equation}\label{eq17k}
  w_t-[v(1-w^2)]_x=0,\qquad v_t-[w(\rt-v^2)]_x=0.
\end{equation}
These are Ovsyannikov equations \cite{ovs-79} derived long ago for internal waves in
two-fluid systems. Characteristic velocities of this system are equal to
\begin{equation}\label{eq18k}
  v_{\pm}=2vw\pm\sqrt{(\rt-v^2)(1-w^2)}.
\end{equation}
Eqs.~(\ref{eq17k}) can be cast to the diagonal Riemann form
\begin{equation}\label{eq19k}
  \frac{\prt r_+}{\prt t}+v_+\frac{\prt r_+}{\prt x}=0,\qquad
  \frac{\prt r_-}{\prt t}+v_-\frac{\prt r_-}{\prt x}=0
\end{equation}
for the Riemann invariants
\begin{equation}\label{eq20k}
  r_{\pm}=vw\pm\sqrt{(\rt-v^2)(1-w^2)}.
\end{equation}
The characteristic velocities are expressed in terms of the Riemann invariants by
simple formulas
\begin{equation}\label{eq21k}
  v_+=\frac12(3r_++r_-),\qquad v_-=\frac12(r_++3r_-).
\end{equation}

The simplest soliton solution of Eqs.~(\ref{eq10}) (with $V_1=V_2=0$) was found in 
Ref.~\cite{qps-16} and it corresponds to zero background relative velocity, $v_0=0$,
and equal background densities of the components, $w_0=0$, far from the soliton.
The general soliton solution with arbitrary values of the background parameters
$v_0,w_0$ was obtained in Ref.~\cite{ikcp-17}. We don't need here the explicit form
of this solution.  Some necessary most important for us
relations can be found without knowledge of the full solution. For example, according
to the old remark of Stokes \cite{stokes}, the exponentially small soliton's tails
$\propto\exp[\pm\kappa(x-Vt)]$, $V$ being the soliton's velocity, and the small-amplitude
harmonic waves $\propto\exp[i(kx-\om t)]$ obey the same linearized system (\ref{eq15k}).
Consequently, the replacement $k\to i\kappa$ transforms the harmonic wave to the 
soliton's tail and the phase velocity $\om/k$ of a harmonic wave transforms to the
soliton's velocity,
\begin{equation}\label{eq22k}
  V=\frac{\om(i\kappa)}{i\kappa}.
\end{equation}
As a result, we obtain the expression for the soliton's velocity
\begin{equation}\label{eq23k}
  V=2vw\pm\sqrt{(\rt-v^2)(1-w^2)-\kappa^2},
\end{equation}
where $v,w$ are the local values of the large-scale wave variables obeying the
dispersionless equations (\ref{eq17k}) and $\kappa$ is the soliton's inverse half-width.

Another relationship
\begin{equation}\label{eq24k}
  \kappa^2=(\rt-v^2)(1-w^2)-(q-vw)^2
\end{equation}
($q$ is a constant parameter) can be obtained using the fact that in completely integrable
system (\ref{eq10}) (with $V_1=V_2=0$) the carrier wave number of a wave packet propagating
along a large-scale wave is only a function of the local values of $v$ and $w$ given by
the formula (see \cite{ks-24b})
\begin{equation}\label{eq25k}
  k^2=(q-vw)^2-(\rt-v^2)(1-w^2),
\end{equation}
where $q$ is an integration constant determined by the values of $k$ at some initial point.
Again the Stokes transformation yields the formula (\ref{eq24k}). Combining Eqs.~(\ref{eq24k})
and (\ref{eq25k}), we obtain (for the choice of the upper sign and supposing  $q-vw>0$)
\begin{equation}\label{eq26k}
  V=\frac{dx}{dt}=vw+q,
\end{equation}
where $x=x(t)$ is the soliton's path.

It is instructive to derive this relation by another method based on the analysis of the
Whitham modulation equations at their soliton limit. It was used in Refs.~\cite{she-18,seh-21}
for consideration of soliton-mean flow interaction and we generalize here this approach.
Periodic solutions of the system (\ref{eq10}) with $V_1=V_2=0$ are parameterized by four
constant parameters $r_i,$ $i=1,2,3,4,$ which in a modulated wave become slow functions
of $x$ and $t$ (see \cite{ikcp-17}). Their slow evolution is governed by the Whitham
modulation equations
\begin{equation}\label{eq27k}
  \frac{\prt r_i}{\prt t}+v_i\frac{\prt r_i}{\prt x}=0,\quad i=1,2,3,4.
\end{equation}
At the soliton limit we have $r_2=r_3$ and these equations reduce to
\begin{equation}\label{eq28k}
  \frac{\prt r_1}{\prt t}+\frac12(3r_1+r_4)\frac{\prt r_1}{\prt x}=0,\qquad
  \frac{\prt r_2}{\prt t}+\frac12(r_1+2r_2+r_4)\frac{\prt r_2}{\prt x}=0,\qquad
  \frac{\prt r_4}{\prt t}+\frac12(r_1+3r_4)\frac{\prt r_4}{\prt x}=0,
\end{equation}
and the phase velocity of the periodic solution becomes the soliton's velocity equal to
\begin{equation}\label{eq29k}
  V=\frac12(r_1+2r_2+r_4).
\end{equation}
Consequently, the soliton's velocity changes along its path according to the equation
\begin{equation}\label{eq30k}
  \frac{dV}{dt}=\frac{\prt V}{\prt t}+V\frac{\prt V}{\prt x}=
  \frac12\left[(r_2-r_1)\frac{\prt r_1}{\prt x}-(r_4-r_2)\frac{\prt r_4}{\prt x}\right].
\end{equation}
The first and third Eqs.~(\ref{eq28k}) coincide with Eqs.~(\ref{eq19k}), that is they
describe evolution of the large-scale wave at the soliton's location and we can identify
$r_1=r_-, r_4=r_+$, where $r_{\pm}$ are defined bt Eqs.~(\ref{eq20k}). Substitution of these
formulas and of $r_2=V-(r_1+r_4)/2=V-vw$ into Eq.~(\ref{eq30k}) yields
\begin{equation}\label{eq31k}
  \frac{dV}{dt}=(1-3w^2)vv_x+(\rt-3v^2)vv_x+V(vw)_x=(vw)_t+V(vw)_x=\frac{d(vw)}{dt}.
\end{equation}
Integration of this equation gives at once Eq.~(\ref{eq26k}).

\section{Hamilton equations}

In derivation of the canonical momentum and Hamiltonian, we still assume that the background wave
obeys the dispersionless equations (\ref{eq17k}) with $\rt=\mathrm{const}$. It is convenient
to introduce a new variable $\vphi$ instead of $\kappa$,
\begin{equation}\label{eq32k}
  \kappa=\sqrt{(\rt-v^2)(1-w^2)}\sin\vphi.
\end{equation}
Then Eq.~(\ref{eq23k}) takes the form
\begin{equation}\label{eq33k}
  V=2vw+\sqrt{(\rt-v^2)(1-w^2)}\cos\vphi
\end{equation}
and substitution of this expression into Eq.~(\ref{eq26k}) gives
\begin{equation}\label{eq34k}
  vw+\sqrt{(\rt-v^2)(1-w^2)}\cos\vphi=q.
\end{equation}
We will need a useful formula which follows from this identity. To obtain it, we differentiate
(\ref{eq34k}) along the soliton's path $x=x(t)$, so that $d/dt=\prt/\prt t+V\prt/\prt x$, and
eliminate $\prt v/\prt t$, $\prt w/\prt t$ with the use of Eqs.~(\ref{eq17k}). Then evident
simplifications yield the necessary formula
\begin{equation}\label{eq35k}
  \frac{d\vphi}{dt}=-\left[\sqrt{(\rt-v^2)(1-w^2)}\right]_x\sin\vphi.
\end{equation}

Now we interpret Eq.~(\ref{eq33k}) as the Hamilton equation
\begin{equation}\label{eq36k}
  \frac{dx}{dt}=2vw+\sqrt{(\rt-v^2)(1-w^2)}\cos\vphi=\left(\frac{\prt H}{\prt p}\right)_x
\end{equation}
and assume that the momentum has the form
\begin{equation}\label{eq37k}
  p=(\rt -v^2)(1-w^2)f(\vphi),
\end{equation}
where $f(\vphi)$ is an unknown function. Integration of Eq.~(\ref{eq36k}) gives
\begin{equation}\label{eq38k}
  H=2vwp+\left[(\rt -v^2)(1-w^2)\right]^{3/2}\int\cos\vphi f'(\vphi)d\vphi.
\end{equation}
To find $f(\vphi)$, we turn to another Hamilton equation
\begin{equation}\label{eq39k}
  \frac{dp}{dt}=-\left(\frac{\prt H}{\prt x}\right)_p.
\end{equation}
Substitution of Eqs.~(\ref{eq37k}) and (\ref{eq38k}) gives after evident simplifications
\begin{equation}\label{eq40k}
  \begin{split}
  &\left[(\rt -v^2)(1-w^2)\right]_x\sqrt{(\rt -v^2)(1-w^2)}f\cos\vphi+(\rt -v^2)(1-w^2)f'\frac{d\vphi}{dt}\\
  &=  -\frac32\sqrt{(\rt -v^2)(1-w^2)}\left[(\rt -v^2)(1-w^2)\right]_x\int\cos\vphi f'd\vphi-
  \left[(\rt -v^2)(1-w^2)\right]^{3/2}\cos\vphi f'\left(\frac{\prt\vphi}{\prt x}\right)_p.
  \end{split}
\end{equation}
Differentiation of Eq.~(\ref{eq37k}) with respect to $x$ for constant $p$ gives
\begin{equation}\label{eq41k}
  \left(\frac{\prt\vphi}{\prt x}\right)_p=-\frac{f\left[(\rt -v^2)(1-w^2)\right]_x}{f'(\rt -v^2)(1-w^2)}.
\end{equation}
We substitute Eqs.~(\ref{eq35k}) and (\ref{eq41k}) into Eq.~(\ref{eq40k}) and arrive at the equation
\begin{equation}\label{eq42k}
  f'\sin\vphi=3\int^{\vphi}\cos\vphi\,f'd\vphi
\end{equation}
which can be solved at once to give
\begin{equation}\label{eq43k}
  f=C(\vphi-\sin\vphi\cos\vphi),
\end{equation}
where $C$ is an integration constant. Its arbitrariness corresponds to the invariance of the
Hamilton equations with respect to multiplication of $H$ and $p$ by the same constant factor.
We choose $C=2$ to simplify some further formulas. Then Eqs.~(\ref{eq37k}) and (\ref{eq38k})
yield after exclusion of $\vphi$ with help of Eq.~(\ref{eq33k}) the following expressions for 
the canonical momentum
\begin{equation}\label{eq16}
  p=2(1-w^2)(\rt-v^2)\arccos\frac{\dot{x}-2vw}{\sqrt{(1-w^2)(\rt-v^2)}}-
  2(\dot{x}-2vw)\sqrt{(1-w^2)(\rt-v^2)-(\dot{x}-2vw)^2},
\end{equation}
and for the Hamiltonian
\begin{equation}\label{eq17}
  H=2vwp+\frac43\left[(1-w^2)(\rt-v^2)-(\dot{x}-2vw)^2\right]^{3/2},
\end{equation}
where $x=x(t)$ is the soliton's trajectory.

It is worth noticing that these canonical momentum and Hamiltonian are related with those
functions in Hamiltonian mechanics of NLS equation solitons \cite{ik-22,kamch-23} by simple
formulas. Indeed, the `shallow water equations'  
\begin{equation}\label{eq13}
  n_t+(nu)_x=0,\qquad u_t+uu_x+n_x=0,
\end{equation}
which represent the dispersionless limit of the NLS equation, can be transformed to the
dispersionless equations (\ref{eq17k}) by means of substitutions
\cite{ovs-79,kamch-23}
\begin{equation}\label{eq14}
  u=2vw,\qquad n=(1-w^2)(\rt-v^2).
\end{equation}
Consequently, these substitutions cast the expressions for the canonical
momentum and Hamiltonian of the NLS soliton to the expressions (\ref{eq16}) and (\ref{eq17}),
respectively. This observation allows us to simplify derivation of the Newton equation
for the magnetic soliton motion.

\section{Newton equation}

In Hamiltonian mechanics, we have to solve Eq.~(\ref{eq16}) with respect to $\dot{x}$
and to substitute this function of $x$ and $p$ into Eq.~(\ref{eq17}). Then the
Hamilton equations (\ref{eq36k}) and (\ref{eq39k}) govern the soliton's motion. 
It is impossible to perform
such an elimination of $\dot{x}$ explicitly, therefore it is more convenient to
pass from the Hamilton equations to the Newton-like equation for the soliton's
acceleration $\ddot{x}$. Actually, this transformation was already put in
practice for the NLS soliton in Ref.~\cite{ik-22} and we can use the formulas
obtained there for writing the Newton equation for the magnetic soliton with help of 
the substitutions (\ref{eq14}). The Newton equation for the NLS soliton reads \cite{ik-22}
\begin{equation}\label{eq49k}
  2\ddot{x}=n_x+(u+\dot{x})u_x+2u_t+\frac{n_t+(nu)_x}{\sqrt{n-(\dot{x}-u)^2}}
  \arccos\frac{\dot{x}-u}{\sqrt{n}}.
\end{equation}
Consequently, the substitutions (\ref{eq14}) yield the Newton equation for the
magnetic soliton in the following form
\begin{equation}\label{eq18}
  \begin{split}
  & 2\ddot{x}=\left[(1-w^2)(\rt-v^2)\right]_x+2(2vw+\dot{x})(vw)_x+4(vw)_t\\
  & +\frac{\left[(1-w^2)(\rt-v^2)\right]_t+\left[2(1-w^2)(\rt-v^2)vw\right]_x}
  {2\sqrt{(1-w^2)(\rt-v^2)-(\dot{x}-2vw)^2}}
  \arccos\frac{\dot{x}-2vw}{\sqrt{(1-w^2)(\rt-v^2)}}.
  \end{split}
\end{equation}
As one can see, in this theory the soliton's dynamic is governed by the gradients of
the background distributions in agreement with our assumption that the soliton's
width is much smaller than the characteristic length at which the background changes,
so the soliton's profile is only determined by the soliton's velocity and the local
values of the background variables. These variables evolve according to the
dispersionless equations obtained from Eqs.~(\ref{eq10}) by neglecting the dispersion
term in the second equation,
\begin{equation}\label{eq19}
   w_t=\left[v(1-w^2)\right]_x,\qquad v_t=\left[w(\rt-v^2)\right]_x+V_{1,x}-V_{2,x}.
\end{equation}
In derivation of expressions for the canonical momentum and the Hamiltonian, we assumed 
that the energy and momentum do not depend on the gradients of the potentials, so we 
used Eqs.~(\ref{eq17k}). Now, in discussion of the large scale dynamics of the
background, we have to take into account the gradients of the potentials $V_{1,x}$
and $V_{2,x}$ as well as slow $x$-dependence of the mean total density in the
variable $\rt=2\delta\rho_0$, where $\rho_0(x)$ is a function of $x$ determined by 
Eq.~(\ref{eq4}). Thus, we can simplify Eq.~(\ref{eq18}) by exclusion of $w_t$ and
$v_t$ with the use of Eqs.~(\ref{eq19}) and after some manipulations we get
\begin{equation}\label{eq20}
  \begin{split}
   \frac{d^2x}{dt^2}=\frac{d(vw)}{dt} +\frac{1+w^2}2\rt_x +w(V_1-V_2)_x
   -\frac{v(1-w^2)(V_1-V_2)_x}
  {\sqrt{(1-w^2)(\rt-v^2)-(\dot{x}-2vw)^2}}
  \arccos\frac{\dot{x}-2vw}{\sqrt{(1-w^2)(\rt-v^2)}}.
  \end{split}
\end{equation}
Let us consider some particular cases.

If $V_1=V_2=0$ and, consequently, $\rt=\mathrm{const}$, this equation reduces to
$d^2x/dt^2=d(vw)/dt$ so its evident integration reproduces to the relationship (\ref{eq26k}).
This equation is very convenient for discussions of the soliton's motion along non-uniform
evolving background waves which are not subject to action of external potentials. For example,
if the soliton moves along a self-similar rarefaction wave evolved from and initial
discontinuity \cite{hamner-11,ckp-16}, then of the Riemann invariants (\ref{eq20k}) is
constant. Let it be $r_-=r_-^0=\mathrm{const}$, then one the other invariant $r_+$ is 
determined by the equation $v_+=x/t$, so these two equations give $r_-+v_+=3vw=x/t+r_-^0$.
Equation
\begin{equation}\label{eq52k}
  \frac{dx}{dt}=\frac13\left(\frac{x}{t}+r_-^0\right)+q=\frac13\frac{x}{t}+q_1,
\end{equation}
where $q_1=q+r_-^0/3$ is determined by the parameter $r_-^0$ of the background flow and by
the initial soliton's velocity at the moment $t=t_0$. This equations is readily integrated 
to give
\begin{equation}\label{eq53k}
  x(t)=\frac32q_1t+Ct^{1/3},
\end{equation}
where the integration constant $C$ is determined by the initial soliton's coordinate $x_0=x(t_0)$.

Equation (\ref{eq20}) greatly simplifies, if there is no relative motion of the components ($v=0$),
\begin{equation}\label{eq21}
  \frac{d^2x}{dt^2}=\frac{1+w^2}2\rt_x+w(V_1-V_2)_x,
\end{equation}
where $\rt(x)=2\delta g\,\rho_0(x)=(\delta g/g)(\mu-V_1(x)-V_2(x))$ (see Eq.~(\ref{eq4})).
If the trap potential is the same for both components, $V_1=V_2=V_{ext}$, then $w=\mathrm{const}$
and Eq.~(\ref{eq21}) reduces to
$$
\frac{d^2x}{dt^2}=-\frac{(1+w^2)\delta g}{g}\frac{dV_{ext}}{dx}.
$$
Here we should remember that in these calculations we used the space variable $x'=\sqrt{2}x$ (see 
text below Eq.~(\ref{eq6})), so returning to the usual coordinate $x$, we write this equation as
\begin{equation}\label{eq55}
  \frac{d^2x}{dt^2}=-\frac{(1+w^2)\delta g}{2g}\frac{dV_{ext}}{dx}.
\end{equation}
Multiplication by $\dot{x}$ and integration yield the integral of motion
\begin{equation}\label{eq21b}
  \frac12\left(\frac{dx}{dt}\right)^2+\frac{(1+w^2)\delta g}{2g}V_{ext}=\mathrm{const}.
\end{equation}
This equation determines the soliton's path along a stationary profile of condensate's density 
$\rho_0(x)$ confined in the potential $V_{ext}(x)$. In a linear potential $V_{ext}=\eta x$ the
soliton moves according to the ``Galileo law''
\begin{equation}\label{eq21c}
  x=x_0-\frac{(1+w^2)\delta g}{4g}t^2.
\end{equation}
In a harmonic trap $V_{ext}=(\om_0^2/2)x^2$
the soliton oscillates with the frequency
\begin{equation}\label{eq57}
  \om=\sqrt{\frac{(1+w^2)\delta g}{2g}}\om_0.
\end{equation}
In case of equal densities of the components ($w=0$), we get
\begin{equation}\label{eq58}
   \om=\sqrt{\frac{\delta g}{2g}}{\om_0}.
\end{equation}
It differs from the frequency $\om=\om_0/\sqrt{2}$ of oscillations of a dark soliton in a
one-component condensate \cite{ba-2000,kp-04} by the factor $\sqrt{\delta g/g}$.

If $V_1=-V_2=V_{ext}(x)$ \cite{brfrp-23}, then the total density is constant, 
$\rho_1+\rho_2=\rho_0=\mathrm{const}$, as well as $\rt=2\delta g\rho_0$, but the partial 
densities are non-uniform and the second Eq.~(\ref{eq19}) yields the distribution of $w$,
\begin{equation}\label{eq59}
  w(x)=-\frac{1}{\delta g\,\rho_0}V_{ext}(x),
\end{equation}
where we have chosen the integration constant in such a way that $w=0$ at the point with
$V_{ext}=0$. In this case Eq.~(\ref{eq21}) reduces with account of the replacement
$x\to\sqrt{2}x$ to
\begin{equation}\label{eq60}
  \frac{d^2x}{dt^2}=-\frac{1}{2\delta g\rho_0}\frac{d(V_{ext})^2}{dx},
\end{equation}
and this equation has an integral of motion
\begin{equation}\label{eq61}
  \frac12\left(\frac{dx}{dt}\right)^2+\frac{1}{2\delta g\rho_0}V_{ext}^2=\mathrm{const}.
\end{equation}
If $V_{ext}=\eta x$, then soliton oscillates with the frequency
\begin{equation}\label{eq62}
  \om=\frac{\eta}{\sqrt{\delta g\,\rho_0}}.
\end{equation}

\section{Conclusion}

The method used in this paper is based on the generalization of Stokes' 
remark~\cite{stokes} that soliton's tails propagate with soliton's velocity and
at the same time satisfy the linearized equations for small-amplitude waves. Therefore, the
soliton's velocity is related with the soliton's inverse half-width by the dispersion relation
continued to the region of complex wave numbers. Validity of this remark is confirmed by
particular soliton solutions of a number of concrete nonlinear wave equations. It remains
also correct for solitons propagating along non-uniform background, as one can see from
the coincidence of the equation for soliton's motion with the corresponding results 
obtained in the soliton limit of the Whitham modulation equations in completely integrable
cases when the Whitham equations are available in the Riemann diagonal form. Extension
of the obtained by this method equations for solitons motion to non-integrable situations 
allows one to reproduce some known results and to obtain the new ones.

This approach is not limited to the problem of solitons motion. Another its interesting 
application is to the quasi-classical method of calculation of velocities of solitons
produced from an intensive initial pulse \cite{kamch-23b}. In completely integrable cases,
it easily reproduces the known results following from the WKB method applied to the
spectral problem associated with the concrete equation in framework of the inverse 
scattering transform method. What is more important, it can be applied to not completely
integrable equations and gives the results which agree very well with the results of
numerical simulations.

At last, we should mention that our approach allowed us to connect the condition of
preservation of Hamiltonian mechanics of propagation of high-frequency wave packets 
along a smooth large-scale dispersionless flow with the quasi-classical limit of the 
Lax pair for equations integrable in framework of AKNS scheme \cite{kamch-23b,ks-24b}.
This observation sheds new light on conditions of complete integrability of nonlinear
wave equations.

One may hope that the suggested method can lead to many other interesting results.

\section*{Acknowledgments}

I am grateful to N.~Pavloff and D.V.~Shaykin for useful discussions.
This research is funded by the research project FFUU-2024-0003 of the Institute of Spectroscopy
of the Russian Academy of Sciences (Sections~2,3) and by the RSF grant number~19-72-30028
(Sections~4,5).

\end{document}